# Work on the Manhattan Project, Subsequent Events, and Little Known Facts Related to its Use

Lawrence S. Bartell

Department of Chemistry, University of Michigan, Ann Arbor, MI 48109

*lbart@umich,edu*

**Abstract:** A personal account of work on the Manhattan Project in Chicago by one of the few remaining survivors of the war-time project is given, illustrating, among other things, how absurd things can happen at a time of great stress and concern.. As is well known, Los Alamos was the site specializing in the physics of the bomb while Chicago emphasized metallurgical and chemical research. Nevertheless, physics played a significant role in Chicago, as well. That is where Fermi constructed the world's first uranium pile under the stands of Stagg field, a site at which this author got seriously irradiated. Some curious events occurring after the bomb was dropped are also related. In addition, at this time of public protest by sincere people who question the ethics of America's dropping of the bomb on innocent civilians, certain facts, obviously unknown to the protesters, are presented which place the bombing in a rather different light.

.

## Work on the Manhattan Project

Late in 1943, when I was in the final semester of my (draft-deferred) study of science, I was invited to the University of Chicago to be interviewed for a job on a secret war project, part of the "Manhattan Project." So I went, suspecting it to be work on some aspect of the atomic bomb--because hints that such a device was possible had appeared in popular magazines. When I arrived, none other than Glenn Seaborg interviewed me. He was later awarded the Nobel Prize for his discovery of plutonium, a new element he had created by irradiating uranium with neutrons. Plutonium is the fissionable material used to power the Nagasaki bomb. After a short interview, Dr. Seaborg asked when I could start. I replied, 'Today, except for one detail. Final exams for my undergraduate degree are going to be held next week."  Seaborg told me to wait a few minutes and went to the telephone.  He came back with a smile and told me to begin right away.  He had arranged to have me excused from the exams.  Actually, this rather alarmed me because I wasn't sure I was even passing Economic Geography, a course I didn't care for and had scarcely paid attention to.  Well, it seemed that the urgency of this project to the war effort and the possible  importance of my contribution to it (rightly or wrongly), was a message that must have been conveyed eloquently by Seaborg (so I did end up passing the worrisome course)

Shortly after I began work I was given a small flask containing one gram of plutonium in solution, being told it was worth (in 2009 dollars) over 10 million dollars.  From this I was to take one million dollars' worth for my own research.  To a 20-year-old kid fresh out of academic classes, this was incredible.  My job, in coordination with the work of several other young scientists, was to test ways of efficiently extracting the small amounts of plutonium present in the heavy uranium slugs from a nuclear reactor.  The plutonium had been produced by irradiating the dominant uranium isotope ($^{238}$U, mass 238 atomic units) with some of the neutrons resulting from the fissions of the nuclei of the rare isotope ($^{235}$U, present to the extent of only 0.7 %), leaving the rest of the neutrons from the fissions to cause other $^{235}$U nuclei to fission, thereby keeping the nuclear chain reaction going. The product sought, plutonium, element 9<u>4</u>, mass 23<u>9</u>, was code named "49." This led to a ridiculous situation.  To help maintain our fitness and morale, the project had a baseball league.  I was first baseman of the "Thompson Commandos." Another team, mostly of Californians, chose the name "49ers."  Project security vetoed that name because it was classified!

As most readers know, the isotope $^{235}$U can itself be made into a bomb, and once the isotope is available in nearly pure form, a uranium bomb can be constructed much more easily than a bomb of plutonium. As a matter of fact, the physicists were so confident that a uranium bomb would work that it was never tested before one was dropped on Hiroshima. Plutonium was a different kettle of fish. Because it contained traces of the spontaneously fissionable isotope of mass 240, which constantly emits neutrons, the bomb would fizzle out prematurely if it were designed the same way as the uranium bomb. A very tricky implosion technique was required to make the system supercritical rapidly enough to avoid having it fizzle before the chain reaction was sufficiently complete to produce a true atomic explosion. Working out this technique proved to be one of the most difficult problems to be solved. Scientists were sufficiently uncertain that a plutonium bomb would even work that they tested one in the New Mexico desert. This test was carried out the month before the atomic bombs were dropped on Japan.

There were many difficulties associated with plutonium, that might raise questions about its production. Of course, the production of plutonium in nuclear reactors is at the expense of the fissile rare isotope $^{235}$U in natural uranium. How to design nuclear reactors was also problematic. Reactors need a "moderator" to slow down the neutrons produced by fissions of $^{235}$U, and the required moderator must be a liquid or solid material containing light atoms that don't absorb neutrons. Ordinary water, $H_2O$, would be ideal except for the fact that hydrogen atoms sometimes absorb a neutron to produce deuterium, "heavy hydrogen," whose chemical symbol is "D." On the other hand, heavy water, $D_2O$, *is* a good moderator but, being a rare component of ordinary water, it is not easy to produce in a pure state.  Graphite, a form of the light element carbon, is also a reasonably good moderator.  Nevertheless, if graphite is used in reactors, it must be of unprecedented purity because graphite, as ordinarily produced, contains traces of boron, an element that soaks up neutrons and quenches a chain reaction before it can get started.  Such graphite was not available until chemists worked out special procedures for its manufacture. I digress here to mention that this trouble with graphite is one of the reasons America far outpaced Germany in the race to obtain a sustained nuclear chain reaction.  The arrogance of German physicists made them suppose they alone had the intellect and talent required to achieve their goal.  They didn't bother to consult chemists and engineers, the very people who made the American project feasible. Tests by German physicists showed that graphite was unsatisfactory as a moderator in a reactor, but the physicists did not realize that the trouble lay in traces of boron.  So they opted for heavy water, a material available only at great expense from one Norwegian hydroelectric plant.  Sabotage prevented the Germans from ever attaining enough heavy water for a nuclear reactor.  Pure graphite can be made in quantity far more cheaply and easily than heavy water.

There were many other difficulties associated with the production of plutonium. The cooling of reactors required a well-filtered river to carry away the heat.  The radiation they produced was a terrible problem to get rid of, and the chemical processes for extracting and purifying plutonium in large amounts could only be done by a tricky remote control.  Before the war, plutonium was unknown and all of its chemical and physical properties had to be learned very quickly.  Pure plutonium has a complex number of phases whose properties render it unsatisfactory for the construction of a bomb.  The metal had to be alloyed with another special metal to make it behave satisfactorily.  Therefore, the purification and the special metallurgy to make a suitable bomb once the purified plutonium was extracted was a ticklish business. Why, then, were the costly reactors ever designed and operated, and why was the enormous plutonium effort ever undertaken in the first place?  Why weren't bombs simply made directly from uranium?  The answer is that the uranium isotope 235, the isotope capable of yielding an explosive chain reaction, had to be enriched from the natural uranium before a uranium bomb was possible.  Since the chemical properties of the light and heavy isotopes of uranium are virtually identical, no chemical process conceived of at that time could be used to achieve the separation.  Physical methods would be needed, such as thermal diffusion, a very inefficient and expensive procedure, or mass spectroscopy using giant mass spectrometers  In fact, both of these methods were used.  But they produced the lighter isotope in such a slow trickle that by the time of Hiroshima there was only enough for a single bomb.  A procedure involving centrifuging has since been shown to be superior to the methods originally used but preliminary tests of it in war-time had been unsuccessful.  On the other hand, troublesome though it was to construct and operate reactors, and nasty though the chemical processes were that extracted and purified the plutonium, nevertheless, plutonium could still be produced at a much faster rate than highly enriched uranium 235. The reason is that chemical separations can be accomplished MUCH more efficiently than physical separations.  That is why the plutonium route was pursued.  That is why, as a chemist, I had a wartime job in Chicago.

Separating traces of plutonium quantitatively from amounts of uranium hundreds of times greater was tough enough, but decontaminating it from the fiendishly radioactive fission products from the reactor was even worse.  Every day for about a year I treated "hot" (radioactive) solutions with various chemicals, precipitated, filtered, redissolved, pipetted, and assayed measured aliquots by various techniques.  Of course, all of us wore rubber gloves to prevent direct contact with the nasty stuff.  Still, every time we left the building, for lunch, dinner, or bed, we had to put our hands into a radiation counter.  If the count was low enough, we were permitted to leave.  If it wasn't, bells rang and lights flashed.  And despite my best precautions, the confounded bells ALWAYS rang and the lights ALWAYS flashed when it was my turn.  That meant I had to subject my hands to a series of "oxidation-reduction" cycles until my hands passed the inspection.  These cycles consisted of rubbing wet potassium permanganate crystals all over my hands until my skin turned black, then rubbing them with wet sodium bisulfite crystals until they were bleached white.  I usually had to go through a number of these cycles before I passed.  It is astonishing just how much abuse from harsh chemicals the skin can tolerate.  Before I started work on the project I had been fond of potato chips and salted nuts, but after living with my contaminated hands for awhile, I couldn't bear the thought of eating anything with my fingers.

This contact with radioactivity was more casual back in those wartime crash-program days than it is now.  One day I was given the "honor" of being the first to use the new "hot lab" that hadn't quite been finished (a euphemism for it being totally unready).  This laboratory had been set up under the west stands of the football field, the very site where Fermi had constructed an "atomic pile" which generated the world's first sustained nuclear chain reaction   My task was to centrifuge a large quantity of hot material.  When I finished and returned to my research building, the moment I stepped through the door, the bells rang and the lights flashed on the hand counter some fifty feet away.  This had never happened before!  I was stripped to the skin by the health physics people and washed down, and every orifice was swabbed out.  My clothes were so contaminated that they had to be thrown away.  I've forgotten how I got home that evening.  I never was compensated for my lost clothes.

One often reads today that during the war the dangers of radiation were not recognized.  That is untrue. It had been discovered long before the war that women who licked brushes dipped into radioactive solutions and painted radium on the hands of watches, developed "phossy jaws," a horrible disease, and the dangers of excessive exposure to X radiation had also been studied extensively.  Therefore, we young men were apprehensive about the risks of sterilization or sickness we might be subjected to.  To ease our fears, our leaders told us during one group meeting that male rabbits had been put into a nuclear reactor and irradiated far more heavily than we ever would be.  When they were taken out, they looked a bit worse for wear at first but had no trouble doing what rabbits are so well known for--reproducing themselves.

During wartime, when rapid results are often more important than economy, silly things can happen.  I remember how a door at the end of my hall was kept propped open by a large, shiny bar of aluminum.  At least we all supposed it was aluminum, an inexpensive material.  One day, however, I idly reached down and picked the bar up.  To my astonishment, this crude doorstop was extremely heavy.  It was pure platinum and worth a fortune!

We were kept abreast of progress on the project.  There was a great sense of urgency because it was known that the Germans also had an atomic project and had a head start on us because the committee appointed by Roosevelt had dithered and dallied for a long time until the British goaded us into action. Especially worrisome was the fact that the German project was led by Heisenberg, one of the most brilliant physicists in the world. However disturbing it might seem to work on something as horrible as an atomic weapon, it was incomparably more horrifying to imagine the consequences if Hitler got the weapon first! Therefore, we worked as hard as we could, sometimes for 24 hours at a stretch. Since the project was so secret, draft boards drafted a large fraction of the young men working at the University of Chicago because our research directors couldn't reveal just how essential to the war effort the men were. So, as soon as the men were drafted from the project into the military service, they were taken into the Corps of Engineers as enlisted men and sent right back to the University of Chicago. This, of course, caused some ridiculous and hilarious situations because in the military, when you have enlisted men, you have to have officers to tell the enlisted men what to do. The trouble was that the enlisted men, all scientists, had "Q Clearance" (access to the secrets) but the officers didn't. The officers hadn't the foggiest notion of what the whole thing was about, and sometimes their curiosity made them try to "pull rank" and demand to be told what was going on. The enlisted men took delight in refusing to reply, infuriating the officers. Actually, the enlisted men, having been sworn to secrecy, had no choice.

When I was a university student, I became so fascinated by the science of radioactive substances that I decided to become a radiochemist. Surely radiochemistry was the most exciting field in science. Then, suddenly, I had entered the field. Well, a year of monotonous work pipetting, precipitating, centrifuging, filtering, and dissolving wickedly radioactive materials, getting thoroughly contaminated day after day after day, removed any aura of mystery or romance.  After a year I had had more than enough radiochemistry for a whole lifetime.  Therefore, it was almost a relief when I finally received my "Greetings from the President of the United States" ordering me to report for military duty.  It turned out that I was the very last man at the project in Chicago to be drafted.  Since the research on methods to separate and purify plutonium from uranium had been successfully completed by then (January 1945) I was the first draftee not to be inducted into the Corps of Engineers and returned to Chicago.  So I chose the Navy instead of the Army because muddy trenches held little fascination.  I became a radio technician (Navy RT) and was slated to be in the first wave of the invasion of Japan. Hiroshima and Nagasaki (and the rheumatic fever I contracted *while on heavy work detail in* a naval hospital which I entered only with scarlet fever) made that unnecessary.

Did I feel guilty for having helped to develop the process that made the Nagasaki bomb possible?  No!  I explain why in detail, in the postscript.

**A Postwar Adventure**

How it came to pass that I got arrested as a possible Russian spy in January 1946 at Trinity, the site of the July 1945 test of the world's first atomic bomb to be exploded (a bomb of plutonium, extracted from uranium by "my process") is another story.   I'll begin it by mentioning that in January, 2000, my wife and I traveled to the southwest to escape Michigan's miserable winter and to visit various places, including Los Alamos. Since the visit was well before the Government accidental burn of several hundred houses in Los Alamos, the place was very attractive, much different from the Spartan village I had first seen in 1946.  With its welcoming citizens, pleasant restaurants, and interesting atomic museum, it contrasted starkly with the Los Alamos that had greeted me with hostility in 1946. The trouble a half-century ago really was my fault. On about Halloween, 1945, when I was discharged from a Naval Hospital (with a 100% disability, and I'm still here!!!) I was advised to spend the winter in the southwest to recuperate. During the Christmas holidays in Ann Arbor I met my old friend Paul Barker, a physicist who was taking a Christmas break from his position in the Los Alamos National Laboratory. Since he needed a ride back to Los Alamos and I had acquired a '37 Ford Coupe to drive myself to the southwest, I offered to take him. On the way to New Mexico we began to think it would be fun to visit "Trinity," the site of the world's first atomic explosion, the site where the plutonium implosion bomb had been tested, as discussed above. Although Paul had a "Q clearance" at Los Alamos (as I had had in the project at Chicago in 1944) he was not high enough in the ranks to have been invited to witness the test. So the closer we got to New Mexico, the more enthusiastic we got about trying to see the test site. We only had the vaguest idea of where Trinity was, so as we got closer to where we guessed the site might be, we stopped and asked people if they knew where it was. They would point and tell us stories about how a cow turned white overnight and how a blind person saw a flash. Confirming the latter claim, one of the exhibits at the Los Alamos Atomic Museum told the story about how a blind person saw a flash. This vague pointing to the presumed site of the explosion led us off the main highway onto a narrow dirt road. We weren't at all sure we were on the right road and at various forks in the road, we became even less certain when we had to chose which branch to take. Finally we came to a huge sign "US GOVERNMENT PROPERTY. NO TRESPASSING." Finally! We no longer doubted we were on the right road. So we kept on for a mile or so until we were stopped at a road block where the military police weren't at all pleased to see us.  We were told to get out and to get out fast.  So we started back, then saw a small road off to the north shielded from view of the military police by fairly tall desert scrub.  After all, who could blame us for trying it?  The road might be a shortcut to the main road we needed to get to.  We drove along, climbed a shallow hill, and when we got over the top, THERE IT WAS BEFORE US!  A sea of green glass maybe a half mile across, surrounded by red desert mud.  Chemists will understand that the heat and radiation from the blast reduced the red ferric iron coloring the mud to green ferrous iron, and a glass (now known as "trinitite") formed when the melted surface layer froze.  No crater had been scooped out because the bomb had been exploded from the top of a very tall tower, a tower that had been vaporized except for vestiges of steel feet.  The bomb did compress the desert floor into a shallow depression.

Our exhilaration at our successful adventure soon turned to concern when a mounted military policeman came galloping from the other side of the site to arrest us.  We were taken to a tent where no one was willing to speak to us.  We waited for the better part of an hour while furious telephoning was going on.  Eventually an army truck drove up and a burly sergeant frisked us for weapons, then threw us into the back of the truck.  We sat on the cold steel (it was freezing and the cargo section of the truck was unheated) and no cushions or blankets had been provided to make the 15 minute ride more comfortable.  Even though our captors took no interest in our personal welfare, we were still elated at what we had seen.  When we arrived at the nearest military base, we were taken immediately to the commanding officer.  He was furious.  "Do you know that you are the first unauthorized persons to enter the site?"  (Our bosoms swelled with pride!)  "Your vehicle is the first unauthorized vehicle to reach the site" (Now I was very proud of my little '37 Ford.)  "Didn't you see that sign 'US GOVERNMENT PROPERTY.  NO TRESPASSING?"  I looked uneasily at Paul and he looked at me.  The CO went on, "NO TRESPASSING!  TRESPASSERS WILL BE PERSECUTED!"  This was so funny I burst out laughing, mentioning that we might be prosecuted but not persecuted. Paul looked as if he didn't want to know me. But it seemed to have been the right thing to do - - even if embarrassing the CO might temporarily have exacerbated the situation - - because it became clear that we were just a couple of smart-aleck 22 year old kids. Kids, moreover, who had or who had had Q clearance and were unlikely to be spies. So finally we were driven back to my car and an army vehicle with a cannon purposely aimed right at us followed us until we were out of the site. From there we drove to Los Alamos where we were met with anything but kind hospitality. We were stopped at the gate and kept in custody until they had finished asking project members whether they knew of any Russian connections involving us. None of this could dampen our exhilaration, though. Our adventure had been a total success.

**Postscript A:**

The following paragraphs are concerned with the events involving the dropping of the atomic bombs. Nina Byers has written an excellent account of the political aspects involved in the use of the bomb.[1] In the present account, alternative considerations and personal views will be discussed.

Many stories have been told about the agonizing guilt felt by many who worked on the atomic bomb. Did I feel guilty for having helped to develop the process that destroyed Nagasaki? Not for a moment. As far as Hiroshima and Nagasaki are concerned, of course it is horrible that hundreds of thousands of innocent people were maimed or killed. Anyone who is remotely thoughtful knows that war is hell. But the bombing of Hiroshima and Nagasaki did end the carnage which had already cost many millions of lives. Not everyone knows that the Japanese military was prepared to sacrifice 20 million additional Japanese lives to turn back the anticipated invasion.

Without the atomic bombs, Emperor Hirohito could never have made his public
announcement to surrender (via a recording  because, since he was considered to be a
God, it was inappropriate for him to speak directly to the people).  Without the bombs,
then, he could never have persuaded the military leaders to accept the bitter admission of
failure in surrendering. Unknown to the military, which could not bear the shame of
surrender and, therefore was scheming to subvert Hirohito's actions, Hirohito made *two*
recordings of his speech to the public announcing his decision to surrender.  Both were
hidden, but one was hidden in a particularly obscure place because Hirohito and his close
associates feared the military leaders would try to confiscate the known recording and
prevent the announcement.   That this fear was not unfounded is confirmed by the actions
of a group of military officers who actually surrounded Hirohito's palace and put
Hirohito under house arrest to prevent him from announcing the surrender of Japan.
They also searched his palace to find the recording.  At the last moment a higher military
authority crushed this coup.

I mention additional facts ignored when the inhumanity of Hiroshima/Nagasaki
bombings is so often broadcast to the world. The bombing of Tokyo caused just as much

Not everyone knows that Japan as well as the US had an atomic bomb project,
one that was led by the brilliant physicist Nishina.  Who doubts that Japan would have
used the bomb in retaliation if they had had it?   After the Hiroshima bombing, the
military told the Japanese scientists they would be given three months to produce the
bomb for Japan.  Of course, that was an absurd order because the project had not got very
far in view of the low priority given the scientists in their attempts to obtain the electronic
and other materials needed for such an undertaking.

I mention additional facts ignored when the inhumanity of Hiroshima/Nagasaki
bombings is so often broadcast to the world. The bombing of Tokyo caused just as much misery and cost just as many lives as the atomic bombs.  Why are there no accusations about that? But the destruction of Tokyo and other industrial cities required thousands of bombers, not just one.  And why, when America is castigated for the criminal inhumanity of the Hiroshima/Nagasaki raids is it never mentioned that the Japanese citizens and their Emperor Hirohito share the blame for tolerating the barbaric behavior of their military forces--whose soldiers, among other atrocities,  <u>wantonly slaughtered far more innocent Chinese civilians in one city alone, Nanjing, than were killed by the two atomic bombs together</u>?  The soldiers did this for sport, to "amuse" the troops and to inure them to killing.  This and similar events had led to the American oil embargo which precipitated Pearl Harbor and our entry into WW II.  Nagasaki brought a quick end to such excesses, and to the war. What is important to note is that the atomic bomb, in ending the conflict, had the effect of saving millions of lives, probably including far more Japanese than Allied lives.  And as far as the killing of innocent Japanese civilians in Hiroshima and Nagasaki is concerned, millions of Allied lives had been lost, too, including service men who had been innocent civilians before being ordered to go to war.  In the Japanese conflict it was the attack on Pearl Harbor, not Hiroshima, which sent them to their deaths.

Perhaps even more important than its role in ending the war, the atomic bomb made the consideration of another World War unthinkable.  The world has gone for the longest period in over a century without a global conflict.  Even politicians, who are not known for their forbearance when international friction approaches the ignition point, are aware of the appallingly destructive power of the Hiroshima bomb.  If they had never known by a real example what horrendous destruction the A-bomb can inflict, but had only been told by scientists what their current stockpile might do, in principle, it is certain that governments would have used it by now in some postwar situation. Even Khrushchev, when goaded by Castro to drop a nuclear bomb on Washington during the Cuban missile crisis, knew that such an action would be utter folly because the consequences would be too terrible to contemplate. Now that H-bombs are literally thousands of times more destructive than the Hiroshima bomb, even politicians realize that atomic warfare must be avoided at all costs. So, horrible though Hiroshima was, its legacy has been to save many, many more lives than it cost. That is why I have never felt pangs of guilt over my role in helping to produce plutonium.

**Postscript B:**

It is of interest to include a comment from a colleague in a former "Iron Curtain" country, Bulgaria, a very able scientist I've met and with whom I correspond frequently. When I learned he had spent some time in Hiroshima after the war, I wondered whether it would spoil our relationship if he knew my role in the Manhattan Project and what I had written about it. Since I like to meet problems head on, I sent him my article. He wrote back:

".....though you may be surprised to know it, my view on the bombing does not differ from yours at all, The controversy is between the humanitarian and the historical points of view..........

Who could deny that the bombing was a personal disaster for so many human beings ...........

From the historical angle, however, personal tragedy means nothing and, hence, the positive role of the Hiroshima-Nagasaki bombing is obvious, at least to me. No doubt, this was the strongest possible deterrent for all war-minded politicians. It is even believed that the bombing saved Japan from Russian occupation and even if only this is true, *I can assure you that it is a sufficient justification, because I know only too very well what it means to live under communist rule."*

[1]  Byers, Nina. "Physicists and the 1945 Decision to Drop the Bomb." arXiv.org>physics>arXix:physics/0210058v1.